\documentclass[pra,preprint,showkeys,nofootinbib,superscriptaddress]{revtex4}%
\usepackage{amssymb}%
\usepackage{graphicx}
\usepackage{amsfonts}
\usepackage{amsmath}
\usepackage{graphicx}
\usepackage{dcolumn}
\usepackage{epsfig}
\usepackage{color}
\usepackage{hyperref}
\usepackage{hyphenat}
\usepackage{bm}
\usepackage{float}
\usepackage{amsthm}
\usepackage{enumitem}
\usepackage[utf8]{inputenc}
\usepackage[T1]{fontenc}
\usepackage{color}
\usepackage{url}
\newcommand{\be}{\begin{equation}}
\newcommand{\beast}{\begin{equation*}}
\newcommand{\ee}{\end{equation}}
\newcommand{\eeast}{\end{equation*}}
\newcommand{\br}{\begin{eqnarray}}
\newcommand{\brast}{\begin{eqnarray*}}
\newcommand{\er}{\end{eqnarray}}
\newcommand{\erast}{\end{eqnarray*}}
\newcommand{\bse}{\begin{subequations}}
\newcommand{\ese}{\end{subequations}}

\newcommand{\bd}{\begin{displaymath}}
\newcommand{\ed}{\end{displaymath}}

\newcommand{\bfig}{\begin{figure}}
\newcommand{\efig}{\end{figure}}

\providecommand{\U}[1]{\protect\rule{.1in}{.1in}}

\begin{document}
\interfootnotelinepenalty=10000
\title{The delimiting/frontier lines of the constituents of matter\footnote{This manuscript is based on 
an article that was published in Brazilian portuguese language in the journal 
\emph{Revista Brasileira de Ensino de Física}, 2018, \textbf{41}, 
e20180160. \color{blue}{\url{http://dx.doi.org/10.1590/1806-9126-rbef-2018-0160}.}}}
\author{Diógenes Galetti}
\affiliation{Instituto de F\'{\i}sica Te\'{o}rica, Universidade Estadual Paulista (UNESP), 
São Paulo, SP, Brasil} \email{diogenes.galetti@unesp.br}
\author{Salomon S. Mizrahi} 
\affiliation{Departamento de F\'{\i}sica, CCET, Universidade Federal de S\~{a}o Carlos, 
S\~{a}o Carlos, SP, Brasil} \email{salomon@df.ufscar.br} 
\begin{abstract}
Looking at the \emph{chart of nuclides} displayed at the URL of the International Atomic Energy 
Agency (IAEA) \cite{IAEA} --  that contains all the known nuclides, the natural and those produced 
artificially in labs -- one verifies the existence of two, not quite regular, \emph{delimiting lines} 
between which dwell all the nuclides constituting matter. These lines are established by the highly 
unstable radionuclides located the most far away from those in the central locus, the \emph{valley of 
stability}. Here, making use of the ``old'' semi-empirical mass formula for stable nuclides together 
with the energy-time uncertainty relation of quantum mechanics, by a simple calculation we show that 
the obtained \emph{frontier lines}, for proton and neutron excesses, present an appreciable agreement 
with the delimiting lines. For the sake of presenting a somewhat comprehensive panorama of the matter 
in our Universe and their relation with the frontier lines, we narrate, in brief, what is currently 
known about the astrophysical nucleogenesis processes.
\end{abstract}
\date{\today}
\startpage{1}
\endpage{40}
\maketitle
\tableofcontents
%
\section{Introduction}
%
At the present stage of our acquaintance with the nature of matter that makes the 
Universe\footnote{Here we are not going to discuss dark matter and dark energy.}, one 
knows\footnote{We adopt the concept of ``established or consolidated knowledge'' according to the 
understanding of the biochemist Max Perutz \cite{PERUTZ1994})} that it is composed of atoms and 
subatomic particles, the atoms are made of a cloud of electrically charged particles, the electrons 
(by convention the charge is negative), that surrounds a massive core, the \emph{nucleus}. The nucleus 
is constituted by two kinds of particles: the protons, which have a positive electric charge, and the 
chargeless neutrons. An electrically neutral atom has the same number of electrons as its number of 
protons; an atom having a different number of electrons than of protons is said to be ionized. A 
chemical element is characterized by the number of protons, $Z$, in its nucleus, and not by the number 
of electrons in its cloud, although the chemical reactions involve, essentially, the Coulomb interaction 
between the electronic clouds of the atoms that may form more complex structures, as molecules and 
crystalline arrays. In a more \emph{latu sensu} definition of matter one should also include the quantum 
description of the electromagnetic radiation that under specific circumstances behaves as being composed 
by fundamental particles called photons. These particles differ, essentially, from one another by the 
amount of energy they carry. A photon in the microwave electromagnetic spectrum carries less energy than 
one in the visible region, which, by its turn, will carry less energy than one in the ultraviolet range. 
At higher energies, a photon in the X-ray spectrum carries less energy than one $\gamma$ photon. From now 
on we will focus on the atoms whose nuclei belong to different species. For the history of the nuclear 
physics origins see, for example, the article by C. Weiner \cite{WEINER1972}.

An atomic nucleus is characterized and identified by its \emph{mass number} $A=Z+N$, where $N$ is its 
number of neutrons. The mass of a neutron is slightly higher than the mass of a proton, and, except for 
its electric charge, they have been formally considered being a single kind of particle, called 
\emph{nucleon}, that can be found out in one of two mutually exclusive states, one having an electric 
charge and the other being devoid of it; as so, a nucleus contains $A$ nucleons. An atom whose nucleus 
have a mass number $A$ and a specific proton number $Z$ is called \emph{nuclide}, a term coined by 
T. P. Kohman in 1947 \cite{KOHMAN1947}. Nuclides having a fixed number of protons but differing from 
each other by their number of neutrons are known as \emph{isotopes} of the chemical element $X$, and 
are represented symbolically as $^{A}_{Z}X_{N}$. For instance, some isotopes of carbon are 
$^{12}_{\, \, 6}C$, $^{13}_{\, \, 6}C$, $^{14}_{\, \, 6}C$ and as the neutron number is $N=A-Z$ the 
subscript $N$ is commonly omitted.

The nuclides are classified into two categories: the stable ones and those that undergo transformations, 
the radioactive or \emph{radionuclides}. A stable nucleus does not transform spontaneously into another 
one, it keeps its identity \emph{ad infinitum}, as long as it does not interact -- by \emph{weak} or 
\emph{nuclear} forces -- with other nuclides, particles or eletromagnetic radiation. On the other hand, 
a radioactive nucleus $^{A}_{Z}X$ undergoes a \emph{spontaneous} transformation decaying into another one 
($^{A_1}_{Z_1}Y$ or $^{A_2}_{Z_2}W$) or, most rarely, it fissions (it breaks, predominantly, into two other 
unbalanced mass nuclei, $A_1 \neq A_2$,). The decay of a nucleus occurs because it follows a natural law: 
a system left for itself will always run toward a configuration state of lower energy, getting rid of any 
energy excess, at the expense of loosing its identity and even disintegrating. Concerning the stable nuclei, 
they don't have any excess of energy to to get rid. It is known, up to the present year (2018), that there 
exist about 252 different kinds of stable nuclides along with other 34 radionuclides whose half-life 
time\footnote{Half-life time is the time it takes for a number of radionuclides of some kind, within a 
sample, to be reduced to half.} is quite long, thus still existing since the formation of Earth (admitted 
that it occurred some $4.5\ Gy$ ago) and considered to be primordial nuclides.

The radioactive nuclides transmute by: (a) emitting heavy particles out of the nucleus, (b) producing and 
ejecting light particles that did not exist previously, (c) capture of an electron from the atomic cloud 
or, (d) fission, which is a rare event. In case (a) a nucleus decays emitting one or two nucleons, or a 
nucleus of smaller mass; in this last case it is more likely to emit a nucleus of helium, $^4_2He$ (also 
known as an $\alpha$ particle). This process, known as \emph{alpha decay}, or $\alpha$ decay, is formally 
represented as
\be
_{Z}^{A}X\longrightarrow ~_{Z-2}^{A-4}Y+ ~ \alpha \ .
\label{alfadec}
\ee
The total mass number $A$ as well as the atomic number $Z$ are the same before and after a decay, 
implying that the number of nucleons and the electric charge are conserved quantities. This is 
also true in the fission process and in nuclear reactions. In the decay (\ref{alfadec}) the energy 
of the nuclide $^A_ZX$ exceeds the sum of the energies of the decay products $^{A-4}_{Z-2}Y$ and 
$^4_2He$, \emph{i.e.} $E\left(  ^A_ZX \right) > E\left(  ^{A-4}_{Z-2}Y \right) + E\left( ^4_2He \right)$, 
noting that this energy excess does not vanish, in fact it is transformed. After its decay the 
difference in energy of a $^A_ZX$ nucleus transforms into the kinetic energies of the products  
$^{A-4}_{Z-2}Y$ and $^4_2He$ that, in the center of mass referential frame, fly apart along a same 
trajectory line but in opposite directions, according to the law of the linear momentum conservation. 
Another form of nuclear radioactive decay is the beta-decay, or $\beta$ decay in short, it is the case 
(b) mentioned above, by which the nuclide $^A_ZX$ transforms into another neighboring one in the chart, 
by either emitting: (b1) an electron, $e^-$, -- the $\beta^-$ decay -- ; (b2) a positron, $e^+$ -- the  
$\beta^+$ decay --; or, case (c), by capturing an electron from the cloud surrounding it, the 
process is known as \emph{electron capture} or EC. The transmutation processes of $\beta^+$ and EC 
compete among themselves. The three processes are written as
\begin{subequations}
\label{a3}
\begin{eqnarray}
_{Z}^{A}X &\longrightarrow &~_{Z+1}^{\; \; A}Y+~e^{-}+\bar{\nu}\quad \left( \beta
^{-}\right)   \label{a3a} \\
_{Z}^{A}X &\longrightarrow &~_{Z-1}^{\; \; A}W+~e^{+}+\nu\quad \left( \beta
^{+}\right)   \label{a3b} \\
_{Z}^{A}X+e^{-} &\longrightarrow &~_{Z-1}^{\; \; A}W+\nu \quad \left( CE\right) .
\label{a3c}
\end{eqnarray}
\end{subequations}
Another kind of particle is emitted concomitantly with the positron and the electron, the 
neutrino ($\nu$), and its antiparticle the anti-neutrino ($\bar{\nu}$); they have a quite tiny mass 
compared to the electron/positron mass and are electrically chargeless. These particles very rarely 
interact with protons and neutrons, albeit being essential to assure that in nuclear processes,  
energy, linear and angular momenta are conserved. The mass of each nuclide in the left hand 
side (LHS) of formulas (\ref{a3}) is greater than the masses of the decay products in the right 
hand side (RHS). Here too, the differences in mass transform into the kinetic energies of the 
products. 

\begin{figure}[hbt]
\centering
\includegraphics[height=5.3in, width=3.6in]{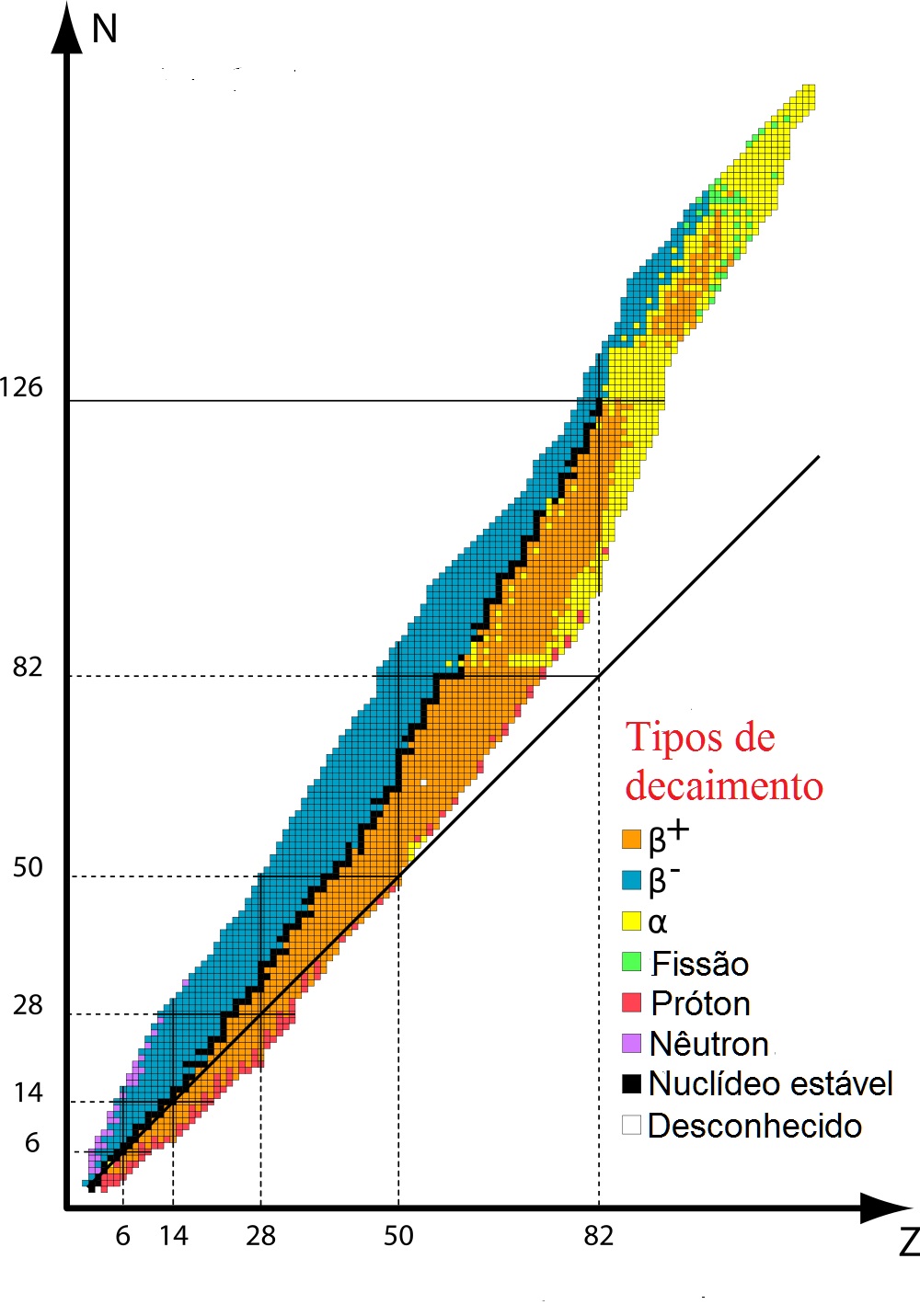}
\caption{\small {Chart of nuclides. The color of each little square denotes a kind of nuclide, 
stable or radioactive, and through which process it decays. The solid line at a $45^\circ$ 
angle corresponds to $N=Z$. By linking the most extreme little squares, corresponding to the 
higher $Z$, or $N$, for a fixed $A$, one gets the \emph{delimiting lines} of the constituents 
of matter. The wiggled central line (in black) links the stable nuclides. As a curiosity, it 
is worth noting the similarity of the paramecia like shape, as illustrated by Otto Müller in 1773, URL 
{\color{blue}{\url{https://en.wikipedia.org/wiki/Paramecium\#/media/File:Muller_paramecium_aurelia.jpg}}}, 
with the nuclides distribution shape.}} 
\label{carta}
\end{figure}

Where and how the nuclides were formed? The scientific approach, developed in the twentieth century,  
gives a sound narrative as it is based on observation, measurement, formal logic and mathematics, 
thus giving rise to the field known as nucleosynthesis or nucleogenesis. The natural nuclides, 
produced in stars, and those obtained in laboratories (using particles and nuclei accelerators) or 
in nuclear reactors, are classified and represented in a 2D chart, with axes $Z\times N$ as shown in 
Fig. \ref{carta}, where each little square stands for one nuclide\footnote{The chart is adapted from 
the URL of the IAEA (International Atomic Energy Agency) \cite{IAEA}.}. This paper presents, initially, 
a short narrative about the origins of the nuclides and then, more emphatically, addresses the question: 
for each $A$, which are the nuclides, with the highest excesses of protons and of neutrons, a mass formula 
predicts? Or, what are the nuclides with $\textrm{max}(N-Z)$ and $\textrm{min}(N-Z)$ that 
form the frontier lines of matter?  

The frontier lines we did calculate are based on a strict formal definition of the concept of 
\emph{drip lines} which are specified as: (a) ``...those that delineate the boundary between bound or 
unbound nuclei.'', (b) ``...the last nucleon that is no longer bound to the lightest or heaviest isotope 
and the nucleus decays on a timescale of strong interactions ($10^{-22}\ s$ or faster)'', (c) ``... the 
boundaries delimiting the zone beyond which atomic nuclei decay by the emission of a proton or neutron''.
A detailed discussion on drip lines can be found, for instance, found in 
\cite{Thoennessen2004,MORRISEY2007,Thoennessen2016}.

This paper is organized as follows: in section \ref{nuc} we present a brief narrative about the nuclides 
synthesis: (1) as it occurs in the core of a star, (2) after the collision and subsequent coalescence of 
neutron stars, and (3) when a star implodes at the final stage of its evolution.
In section \ref{semi} we present and discuss the nuclide semi-empirical mass formula, originally proposed 
by C. F. von Weizsäcker \cite{WEIZS1935} and, independently, by H. A. Bethe and R. F. Bacher \cite{BETHE1936}. 
In subsection \ref{estab} we review the valley of stability (the region which is the \emph{locus} of the 
stable nuclides) that is compared with the \emph{stability line} derived from the mass formula. As an 
aside, in subsection \ref{life} we digress from the core subject of the manuscript to dissert on peculiar 
characteristics of the nuclides that constitute the building blocks of the DNA and RNA 
macromolecules\footnote{DNA and RNA are acronyms for deoxyribonucleic and ribonucleic acids.}. In the chart 
of nuclides one observes that the radionuclides located the most faraway from the valley of stability define 
the delimiting lines that, in section \ref{linhas}, we compare with the frontier lines obtained from  
a nuclide mass formula. Finally, section \ref{conc} contains a summary and conclusions. 
%
\section{Nucleosynthesis}\label{nuc}
%
Starting with hydrogen and up to the iron ($^{26}Fe$) it is currently admitted that the nuclides are 
formed in the core of stars through a sequence of nuclear fusions of lighter ones \cite{BETHE1939,
BURBRIDGE1957}. The study of fusion processes is an essential issue in order to draw a scenario of how 
the chemical elements arose in our Universe and how the stars produce and emit energy in the form of 
electromagnetic radiation and by ejecting highly energetic particles. It is also conceived that nucleons 
and electrons originated from a primordial event, the so-called 
\begin{figure}[hbtp]
\includegraphics[height=2.7in, width=3.2in]{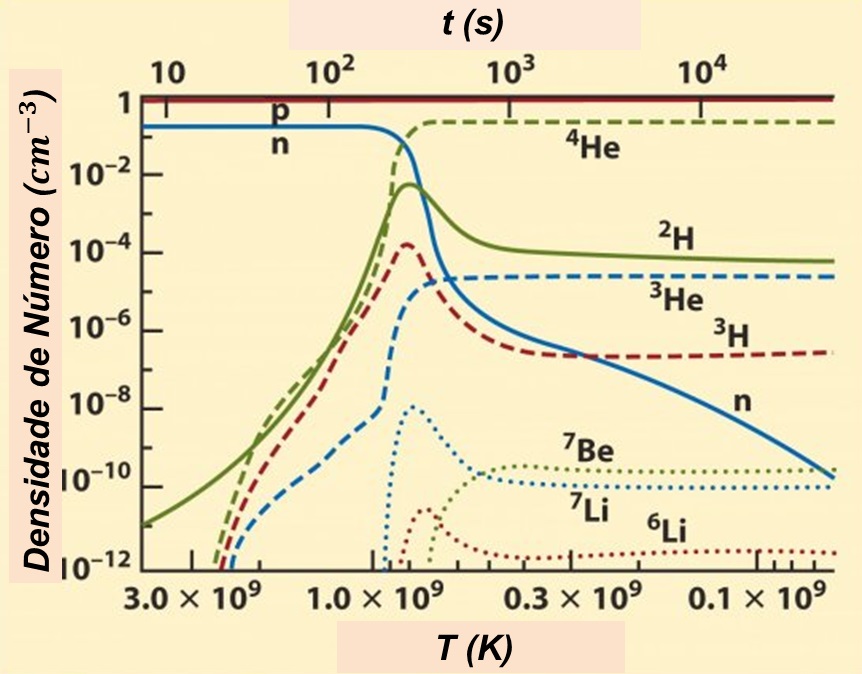}
\caption{\small{Evolution of the relative abundances of elements for the nucleosynthesis 
process in the  primordial Universe; the ordinate stands for the particles number per $cm^3$. 
It is worth noting the relation between the temporal evolution (scale in the top abscissa) 
and the temperature decrease (scale in the bottom abscissa); the formal relation between 
both is $T=10^{10}K/t^{1/2}[s]$, the time $t$ should enter in seconds. Several nuclides 
present a density saturation; see, for example the dashed line for the $_{2}^{4}{He}$. This 
Figure was adapted from the URL {\color{blue}{\url{http://www.astro.ucla.edu/~wright/BBNS.html}}}.}}
\label{BB1}
\end{figure}
\emph{Big Bang}\footnote{This term is attributed to the astrophysicist F. Hoyle \cite{KRAGH2013}. 
It does not mean an ``explosion'', it is about a swift expansion of a primordial hot Universe, as 
deduced from the many astronomical observations and measurements, that confirm the homogeneity and 
isotropy of the distribution of galaxies and clusters of galaxies. As a matter of fact it was observed 
that, in a large-scale, every galaxy and galaxy cluster is moving away from all the others if observed 
in a co-moving reference frame, this behavior is known as the cosmological principle 
\cite{PEEBLES2009}.}, and the evolution in time (or as cooling proceeds) of the relative abundances of 
the primeval elements can be seen in Fig. \ref{BB1}. 

The narrative about the formation of the Universe and the study of the nucleosynthesis of the 
elements is based on the Big Bang paradigm\footnote{We shall not elaborate more on the theory, 
which can be accessed in textbooks specifically devoted to scientific cosmogonies, as for example 
\cite{ROLFS1988,KOLB1990,BASD2005,ILIA2007}, where it is presented in details.}, that 
proposes three main stages of evolution: (i) at times $t\lesssim 0.01\ s$, when the temperature 
was about $T\sim 10 ^{11}K$, an expansion of the Universe occurred\footnote{According to the general 
relativity and modern cosmology the relation between time and temperature of an expanding universe 
is $T \approx 10^{10}K/t^{1/2}[s]$.}, with a great abundance of photons in comparison to the number 
of nucleons; the model estimates a $10^9:1$ ratio. (ii) The second stage is established at 
$t\lesssim 1\ s$ and $T\sim 10 ^{10}K$, the temperature decreases by a factor $10$ in comparison 
with that in the previous stage; an equilibrium ratio of $6:1$ between protons and neutrons is attained. 
(iii) The third stage begins roughly at $t\sim 1$\ min and $T\sim 1  - 3 \times 10 ^{9}K$, the light 
element $^{4}He$ is produced and heavier ones will also be synthesized in sequence, although their 
relative abundances are still relatively quite small as shown in Fig. \ref{BB1}.

Thereafter these stages are over, the Universe continues its expansion and the nucleosynthesis of new 
elements is suspended while the formation of stars and galactic agglomerates begin. As soon as the stars 
take shape, at their core begins the synthesis of new and heavier elements where the pressure is quite 
higher than on its surface. The core is the \emph{locus} where thermonuclear fusion reactions occur and 
chemical elements are produced, from the lighter helium ($A \approx 3-4$) up to iron ($A \approx 56$). 
In complement, the higher gravitational pressure in stars, that are very much heavier than our Sun,  
facilitates the fusion of lighter elements and so heavier ones are produced. The detectable effect of the 
nuclear fusion reactions in stars is their brightness that covers all the electromagnetic spectrum. A 
pedagogical report on stars, their evolution and stability was presented by S. Chandrasekhar 
\cite{CHAND1984}.

In young stars the process of energy production, as a byproduct of nuclear reactions and fusions, 
begins with an initial $3:1$ ratio of hydrogen to helium, which permits a variety of reactions chains. 
They occur in a succession of steps as it was calculated by H. Bethe in a 1939 seminal paper  
\cite{BETHE1939}. For the process known as \emph{pp1} the reaction chain is
\begin{equation}
\left.
\begin{tabular}{ccc}
$p+\ p$ & $\longrightarrow$ & $\ ^{2}H+e^{+}+\nu_{e}$ \\ 
$p+\ ^{2}H$ & $\longrightarrow$ & \ $\ ^{3}He+\gamma $ \\ 
$^{3}He+\ ^{3}He$ & $\longrightarrow$ & $\ ^{4}He+\ p+\ p $ 
\end{tabular}%
\right\}
\label{reacoes1}
\end{equation}%
whose net result is the formation of the $^{4}He$ nucleus plus lighter nuclei, the emission of  
$\gamma$ rays and the release of energy, that go along with the production, in intermediary steps, 
of the lighter nuclei deuterium ($^{2}H$) and helium $^{3}He$. As soon as a high proportion of 
the star hydrogen converts into $^{4}He$ and an enough quantity is produced then the second step of 
the process is launched. That condition is necessary because the gravitational force increases the 
pressure on the core and, consequently, enhancing the probability of collisions among the helium 
atoms themselves, a process that induces novel reaction chains known as \emph{pp2}, 
\begin{equation}
\left.
\begin{array}{ccc}
^{3}He+\ ^{4}He & \longrightarrow  & \ ^{7}Be+\gamma  \\ 
^{7}Be+e^{-} & \longrightarrow & \ ^{7}Li+\nu _{e} \\ 
^{7}Li+p & \longrightarrow & \ ^{4}He+\ ^{4}He \ .%
\end{array}%
\right\}
\label{reacoes2}
\end{equation}%
As an alternative to the second reaction in (\ref{reacoes2}), instead of capturing one electron 
the $^{7}Be$ nucleus captures a proton that finally decays into two stable $^{4}He$ nuclei 
($\alpha$ particles), according to the sequence
\be
\left.
\begin{array}{ccc}
^{7}Be+p & \longrightarrow  & \ ^{8}B+\gamma  \\
^{8}B  & \longrightarrow  & \ ^{8}Be+e^{+}+\nu _{e} \\ 
^{8}Be  &  \longrightarrow& \ ^{4}He+\ ^{4}He \ .
\end{array}%
\right\}
\label{reacoes3}
\ee%
Then the unstable boron-8 ($^{8}B$) nucleus decays into the beryllium-8 ($^{8}Be$), that has 
a quite short half-life time, $\approx 10^{-16}\ s$ (in comparison with other light radioactive 
nuclei), that will undergo a fission into two $\alpha $ particles, without releasing any 
significant amount of energy. Thus the net result is expressed as 
$ p+\ ^{7}Be\longrightarrow \ ^{4}He+\ ^{4}He $. 
 
When $\alpha $ particles are produced in enough quantity in the star core their 
``burning''\footnote{Meaning combustion in chain by nuclear fusion reactions of $\alpha$ particles, 
with the release of energy.} becomes possible. Due to the higher Coulomb barrier this process 
initiates when the temperature is above $10^{8}K$. The eligible stars for this burning are those 
having a mass of the order of, or higher than, $2.5\ M_{\odot }$ ($M_{\odot }$ is the symbol 
for the mass of the Sun) and the nuclear reactions are
\begin{equation}
\left.
\begin{array}{ccc}
^{4}He+\ ^{4}He & \longrightarrow & \ ^{8}Be+\gamma  \\ 
^{8}Be+\ ^{4}He &\longrightarrow & \ ^{12}C+\gamma \ .%
\end{array}%
\right\}
\label{reacoes4}
\end{equation}%

Nevertheless, in a high temperature and pressure environment and preceding its fission the 
$_{4}^{8}Be$ nucleus has a probability to capture an $\alpha $ particle -- an  exothermic reaction -- 
that propitiates the formation of the stable carbon isotope $_{\ 6}^{12}C$ which is an essential 
element for Life on Earth. This process is known as triple-$\alpha$, or $3\alpha$, and it was 
conjectured by the astrophysicist F. Hoyle \cite{HOYLE1954}. 

Thereafter, if enough carbon accumulates, mainly in the core of the star, an additional capture of 
an $\alpha$ particle becomes possible, thus giving rise to the synthesis of the oxygen isotope 
$_{\ 8}^{16}O$, that, by its turn, is used to synthesize the $_{10}^{20}Ne$ isotope, 
\be
\left.
\begin{array}{ccc}
^{12}C+\ ^{4}He &\longrightarrow &\ ^{16}O+\gamma  \\
^{16}O+\ ^{4}He &\longrightarrow &\ ^{20}Ne+\gamma 
\end{array}
\right\}\ .
\ee%
Next, an assortment of nuclear processes comes out allowing the syntheses of new chemical 
elements and heavier nuclides. For example, the carbon-carbon reaction 
\be
^{12}C+\ ^{12}C \longrightarrow \left\{
\begin{array}{ccc}
^{24}Mg &+& \gamma \\
^{23}Na &+& p \\
^{20}Ne &+& \alpha 
\end{array}
\right.
\ee%
requires temperatures about $5-10\times 10^{8}K$ in quite massive stars, 
whereas the oxygen-oxygen reaction %
\be
^{16}O + {^{16}O} \longrightarrow  \left\{
\begin{array}{ccc}
^{32}S &+&\gamma  \\
^{31}P &+&p\ \\
^{31}S &+&n \\
^{28}Si &+&\alpha \ 
\end{array}
\right.
\ee
requires temperatures $T>10^{9}K$. Stars with mass in the range $10-15M_{\odot }$ have the 
thermodynamic condition to gradually fuse heavier nuclei until attaining the element iron. 
Geometrically, the star can be depicted as concentric shells increasing in size, similarly  
to an onion structure, where, in each shell, predominates a specific nuclear fusion process. 
In its core, which remains inert, reside the heavier nuclei, from silicon to iron, and almost 
no more fusions occur as the synthesis of still heavier elements becomes unlikely because the 
pressure is not sufficiently high. When the star composition consists essentially of iron, 
silicon and other neighboring chemical element in the periodic table, at some moment the burning 
by thermonuclear reactions diminishes and in the course of its evolution another process begins: 
due to the gravitational force the star undergoes a contraction and its density increases 
significantly. If the star has accumulated a high quantity of free neutrons, then during its 
contraction a chemical element heavier than iron can be synthesized when an already existing 
nucleus captures a neutron; it is a process of the kind
\be
n +\  _{Z}^{A}X \longrightarrow \ _{\ \ Z}^{A+1}X+\gamma \longrightarrow \left\{ 
\begin{array}{c}
_{Z+1}^{A+1}Y+e^{-}+\bar{\nu}_{e} \\ 
\\ 
_{Z-1}^{A+1}W+e^{+}+\nu _{e}\ .%
\end{array}%
\right. 
\label{addn}
\ee%
After the capture of a neutron by a nucleus $_{Z}^{A}X$ a decay by $\beta ^{+}$ or $\beta ^{-}$ emission 
occurs transforming it into a radionuclide $_{Z+1}^{A+1}Y$ or as $_{Z-1}^{A+1}W$, which, by its turn, 
will decay by cascade until being transformed into a stable nuclide of another chemical element; this 
process develops as a sequence of transmutations. The $\beta ^{\pm }$ decays have an essential role in 
the nucleosynthesis, they may occur within an old imploding neutron star (known as a \emph{supernova} 
\cite{BETHEBROWN1985}), as well as in the coalescence of a binary neutron stars system. A neutron star 
is constituted by a ratio \emph{circa} $10^6$ neutrons/$cm^3$ to every other component, a proton or a 
nucleus. The prevalent theory that explains the nucleosynthesis of elements heavier than iron is based 
on two kinds of reaction processes (\ref{addn}) that need a large flux of neutrons: (a) the \emph{s-process} 
(s for \emph{slow}) is for ``slow capture'' of a neutron, it accounts for the production of nearly half 
the elements beyond iron, occurring in stars in their final evolutionary stage, having a mass between 1 
and 10 $M_\odot$; (b) the \emph{r-process} (r for \emph{rapid}) consists of a ``rapid capture'', it  
occurs in stars having a neutron density around $10^{20}$/cm$^3$, which induces a higher probability 
of capture; see more and thorough details in \cite{COWAN2004, FREBEL2018} as well as in, 
for example, the seminal articles 
\cite{HOYLE1954, BURBRIDGE1957, SEEGER1965, WALLERSTEIN1997}. A third process, the \emph{rp-process}  
for rapid proton capture (hydrogen burning), was proposed in \cite{WALLACE1981, SCHATZ2001}; it may 
occur in proton rich stars whose temperature is above $10^8$K. It is described by the reaction 
\be
p+ _{Z}^{A}X\longrightarrow \ _{Z+1}^{A+1}X+\gamma \longrightarrow \left\{ 
\begin{array}{c}
_{Z+2}^{A+1}Y+e^{-}+\bar{\nu}_{e} \\ 
\\ 
_{\; \; Z}^{A+1}W+e^{+}+\nu _{e}\ .%
\end{array}%
\right. 
\label{addp}
\ee
In this process a transmutation of one element into another of higher atomic number 
($Z \rightarrow Z+1$) proceeds and the released energy is carried out by a $\gamma$ photon.

After presenting this brief narrative on nucleosynthesis and the origin of elements, we advance 
a step further and consider the distribution of the nuclides as pictured in Fig. \ref{carta}.  
In the next sections we discuss the constituents of matter by making using of a nuclide mass 
formula to calculate and draw: (i) the line that links the stable nuclides, and (ii) the 
lines that define the limits for the radionuclides having the highest number of protons and 
neutrons excesses. 
%
\section{Semi-empirical nuclide mass formula}\label{semi}
%
Looking at the chart in Fig. \ref{carta}, or in \cite{IAEA}, one perceives that the number of nuclides  
is finite, $\approx 3300$. A first conclusion to draw out is that nuclei are subjected to a 
saturation tendency in the number of their nucleons: there is no nucleus with arbitrary number of 
protons or neutrons. This limitation is due to: (a) the balance between the attractive, 
strong and short ranged, nuclear force that acts indistinctly between both kinds of nucleons, (b) the 
long range, but weaker, Coulomb repulsion force that acts only between the protons and, (c) the spin 
1/2 degree of freedom of the nucleons that determines their inherent statistics. That saturation finds 
support on the fact that nuclei with $A > 50$ tend to have more neutrons than protons, $N > Z$, which 
is advantageous from the nuclear total energy balance, as a nucleus becomes energetically more tightly 
bounded than one having less neutrons. Except for the ${^1}H$ hydrogen isotope there is no other stable 
nucleus constituted only of protons or only of neutrons\footnote{From the quantum mechanical scattering 
theory of colliding nucleons a negative scattering length is an indication of the non-existence of a 
bound state, and, experimentally, it is found that colliding proton-proton and neutron-neutron show a 
negative scattering length, thus dineutron and diproton cannot have stable structures, see 
\cite{Mitchell2005,Thoennessen2013}. The proton-neutron system has a single bound state: the deuteron.} 

From the chart of nuclides in \cite{IAEA}, we observe that the most unstable nuclei are located 
at the extreme edge to the right as well as to the left of the valley of stability, that is, at the 
borders of the chart one finds nuclei with quite small half-life times, forming the wiggled delimiting 
lines; for this reason they do not actively participate of the ordinary matter constitution but  
contribute for the production, through decay processes, of nuclei that are stable or radioactive 
with non-null relative abundance. A nice exposition on nuclear cartography is found in \cite{SIMPSON2017}.

From a more rigorous point of view, by adopting an approach that takes into account the relevant quantum 
aspects of a many protons and neutrons system, it is found that the calculation of nuclear properties 
involves technical difficulties inherent to the laborious handling of the many-body problem. Nevertheless, 
theories and approximation methods have been developed to include, besides the Coulomb interaction, a variety 
of nuclear forces which permit to calculate and reproduce several observed nuclear properties, although they 
depend on many parameters extracted from empirical data; see, for instance, the textbooks 
\cite{RING1980,ROWE2010}.

In spite of the fact that the results obtained from those methods permit a good quantitative analysis 
of the nuclear systems, other considerations based on simple and intuitive models can also be of great 
value. In this connection, analogies between nuclei and liquid drops -- extensively discussed since the 
1930 decade -- paved the way for the elaboration of the Weissäcker-Bethe-Bacher model 
\cite{WEIZS1935,BETHE1936} from which a nuclide mass formula emerged. From an historical point of view, 
a classical theoretical study, by Lord Rayleigh, about electrified liquid drops showed that when the 
electric charge of a drop exceeds a certain value it may break apart when its shape is put to oscillate  
\cite{rayleigh1,rayleigh2,BW1939,PETERS1980} and \cite{MG2016,helio,wong}. In recent experiments    
\cite{LIAO2017}, electrified ethanol drops were put to rotate and by increasing their angular momentum 
it was observed that they terminate fissioning, whereas non-rotating static drops become unstable and 
undergo fission at the electric charge value predicted by Lord Rayleigh.

The ordinary characteristics of a nucleus, considered as a liquid drop, point to the following general  
properties: (a) the mass and the charge densities present a saturation in the center of the structure; 
(b) the binding energy increases approximately linearly with the mass number $A$; (c) the mean radius is 
proportional to the cubic root of the mass number, \emph{i.e.}, the volume is proportional to $A$.

As such, the nuclide mass formula is a sum of several contributions and the fundamental ones are: 
(a) the masses of its constituents ${ZM_{1^H}+(A-Z)M_{n}}$; (b) the volumetric binding energy ${-a_{v}A}$; 
(c) the surface tension energy ${a_{s}A^{2/3}}$, and (d) the Coulomb interaction energy between the protons 
${a_{c}A^{-1/3}}$. These terms are the analogues of a charged classical liquid drop. The formula is 
complemented with two additional terms: (e) one that takes into account the energy associated with the 
asymmetry between the number of protons and neutrons ${a_{ass}(A-2Z)^{2}A^{-1}}$, and (f) a contribution 
introduced by empirical verification that nuclei having an even number of protons and of neutrons are more 
stable (have lower energy) than neighboring nuclei with different number of constituents; among several 
other proposed formal expressions the term here adopted is ${\eta a_{p}A^{-1/2}}$. 

Here we do not distinguish energy from mass since we utilize the fundamental relation of mass and energy 
equivalence, ${E=mc^{2}}$, and set $c=1$. Hence the expression for a nuclide mass is 
\begin{equation}
 M(A,Z)=ZM_{1_{H}}+(A-Z)M_{n}-a_{v}A+a_{s}A^{2/3}+a_{c}Z^{2}A^{-1/3} \\
+a_{ass}(A-2Z)^{2}A^{-1}+\eta a_{p}A^{-1/2} \; ,
\label{massa1}
\end{equation}
where the parameters must be adjusted to fit the formula to the experimental data, with (\ref{massa1}) 
holding for the greatest number of nuclides. We admitted the values proposed in 1958 by A. H. Wapstra 
\cite{WAPSTRA1958,WAPSTRA1985}: ${a_{v}=15.835\ MeV}$, ${a_{s}=18.33\ MeV}$, ${a_{c}=0.72\ MeV}$,  
${a_{ass}=23.20\ MeV}$, and ${a_{p}=11.20\ MeV}$ with 
\be
\eta =\left\{ 
\begin{array}{ccc}
-1 & \text{for nuclei with } Z\text{ even and } N \text { even} &  \\ 
0 & \text{for nuclei with }Z\text{ even and } N\text{ odd ; or }Z%
\text{ odd and }N\text{ even} &  \\ 
1 & \text{for nuclei with }Z\text{ odd and }N\text{ odd}\ , & 
\end{array}%
\right. 
\label{eta}
\ee
although in the literature one can find other proposed sets of numerical values that differ slightly 
from these. The hydrogen atom, ${^1}H$, and the neutron masses are $M_{1_{H}}=938.783\ MeV$ 
and ${M_{n}=939.565\ MeV}$, respectively.

The worthiness of the mass formula (\ref{massa1}) is due to its aptness to describe several nuclide 
properties, although for light nuclei ($A < 10$) the numerical estimates deviate from the measured 
values because the analogy with liquid drops does not hold well. With the scope of improving this model 
such to introduce, more appropriately, quantum shell effects in nuclei properties as well as the 
geometrical shape of nuclei, additional terms, not present in Eq. (\ref{massa1}), have been proposed, 
such, for instance, those appearing in 
\cite{SEEGER1961,myers,moller,MOLLER-EPJ2016,samanta,basu,lahiri,gangopadhyay}. 
More terms means more empirical parameters that will surely conduct to less deviations from the 
experimental data, but according to our assessment, besides reducing the simplicity of (\ref{massa1}), 
any ``improvement'' will not change expressively our analysis, neither altering the conclusions.  
%
\subsection{Valley of stability and line of stability}\label{estab}
%
Even in its simplest form the mass formula, Eq. (\ref{massa1}), enables the verification of some 
properties about the stable nuclides. As an illustration of the formula performance we draw the line of 
stability and compare it with the data. Indeed -- as is well known and here presented by mere 
completeness reasons \cite{MG2016,helio,wong} --, if we keep fixed the mass number $A$ (that is, 
considering an isobar line, what in Fig. \ref{carta} corresponds to nuclides located on a line 
perpendicular to line $Z = N$), we rewrite Eq. (\ref{massa1}) -- to better exhibit the physical 
content -- as 
\begin{equation}
M\left( A,Z\right) =C_{1}\left( A\right) +C_{2}Z+C_{3}\left( A\right) Z^{2}\
\label{massa2}
\end{equation}
where the coefficients are   
\be
\left.
\begin{array}{lll}
C_{1}\left( A\right) &=&\left( M_{n}-a_{v}+a_{ass}\right) A+a_{\sup
}A^{2/3}+\eta a_{p}A^{-1/2} \\ 
C_{2}&=&M_{1_{H}}-M_{n}-4a_{ass} \\ 
C_{3}\left( A\right) &=&a_{c}A^{-1/3}+4a_{ass}A^{-1}\ ,
\end{array}
\right\}
\label{masscoef}
\ee
from what we observe a quadratic dependence with $Z$ for an isobar line. The point of minimum 
in Eq. (\ref{massa2}) is $Z_{0}\left(A\right) = -C_{2}/\left( 2C_{3}\left( A\right) \right) $ 
(although, in fact, we could have considered the closest integer value of $Z_{0}$) that is expected 
to correspond to the atomic number of the stable nuclide with mass number $A$. In a Cartesian plot 
$A \times Z$ the set composed by all the numbers $\textrm{Int}\left[Z_{0}\left( A\right)\right]$ 
makes\footnote{$\textrm{Int}$ means "take the closest integer of the decimal number in 
$\left[ \bullet \right]$".} the \emph{line of stability}.
\begin{figure}[tbhp]
\centering
\includegraphics[height=3.0in, width=5.0in]{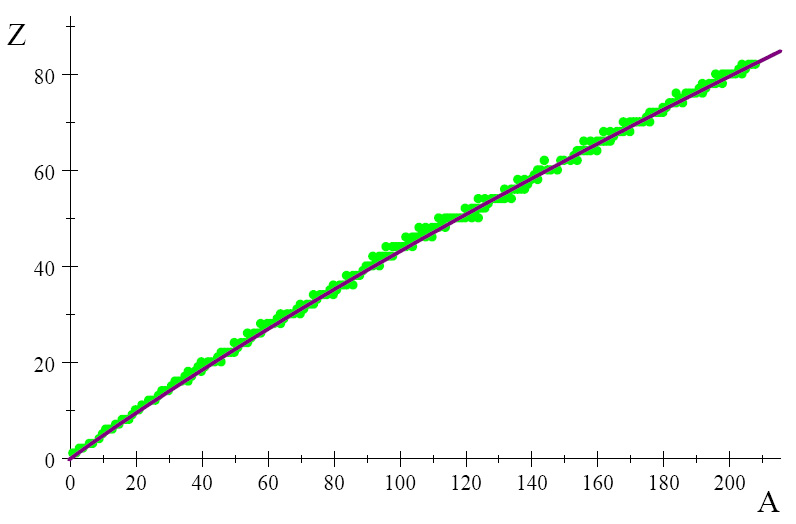}
\caption{\small{Distribution of 252 stable nuclides (green dots, or light gray), composing the 
\emph{valley of stability}; the solid line is the \emph{line of stability} calculated from the 
mass formula.}}
\label{figestaveis}
\end{figure}
In Fig. \ref{figestaveis} the green dots stand for the distribution of 252 stable nuclides composing 
the \emph{valley of stability} whereas the purple line is the calculated line of stability. The 
stable nuclides are surrounded by radionuclides, that will eventually decay to join the nuclides in 
the valley. Up and beyond the valley of stability, new heavy nuclides are unstable, they are usually
identified by the emission of $\alpha$ particles and the limit nuclides are established by the 
identification their fission products \cite{Thoennessen2004,MOLLER-EPJ2016}.  

The remarkable accordance with the empirical data provides a trustful base for the mass 
formula \ref{massa1}, thus it stimulated us to use it to calculate the frontier lines that 
we expected could reproduce well the observed delimiting lines as they show up in the chart of 
nuclides.
%
\subsection{The building blocks of the DNA and RNA macromolecules}\label{life}
%
A quite inspiring exposition of the physics that emerges from the chart of nuclides and its relation 
to Life -- as we know it on Earth -- was elegantly and clearly presented by the nuclear physicist G. Marx in 
an article entitled \emph{Life in the nuclear valley} \cite{MARX2001}. As a complementary discussion on this 
theme we here go further and allude to a fact related to Life and the elements that compose the DNA and RNA 
macromolecules. Since the 1950 decade \cite{KARP2010} it is recognized that they are the material building 
blocks of a molecular structure that contains the necessary information for the codification and production 
of proteins\footnote{Proteins are also large macromolecules but, differently of DNA/RNA, they are not 
related to information storing, each kind have its own specificity, performing a vast array of functions that 
are essential to sustain Life \cite{KARP2010}.} within in biological cells. The DNA/RNA are composed of five 
light chemical elements -- hydrogen ($_1H$), carbon ($_6C$), nitrogen ($_7N$), oxygen ($_8O$), and phosphorus 
($_{15}P$) -- that are structured in a quite special array, the celebrated double helix \cite{KARP2010}. 
Each one of these chemical elements has several natural isotopes that \emph{are not naturally 
radioactive}\footnote{In particular, the radioactive isotope $^{14}C$ is produced in the Earth upper atmosphere 
and it is present on lands and seas at, roughly, a proportion $1:10^{12}$ to all carbon isotopes, so its natural 
relative abundance is admitted to be zero.}, meaning that the sum of the relative isotopic abundances is 
$100 \%$; see Table \ref{dna}. Nonetheless, radioactive isotopes of each of these elements can be produced 
artificially in nuclear reactors. 

The atoms that constitute a DNA/RNA macromolecule are chemically bonded, meaning that there is no 
hindrance for having all the natural isotopes of the five elements in the same proportion as their 
respective relative abundance. These macromolecules are quite stable and resistant against mutations 
(biologically, there exist error corrections mechanisms), however several mutations, when they occur, 
might be quite deleterious to the cells, expressly when induced by external radioactivity, as $\beta, 
\gamma$ and neutron radiation. 
\begin{table}[tbhp]
\centering
\begin{tabular}{|c|c||c|c||c|c||c|c|c||c|}
\hline 
%
$_{1}^{1}$H & $_{1}^{2}$H & $_{\; \; 6}^{12}$C & $_{\; \; 6}^{13}$C & $_{\; \; 7}^{14}$N & $%
_{\; \; 7}^{15}$N & $_{\; \; 8}^{16}$O & $_{\; \; 8}^{17}$O & $_{\; \; 8}^{18}$O & $_{15}^{31}$P \\  \hline 
99.885 & 0.015 & 98.89 & 1.11 & 99.634 & 0.366 & 99.762 & 0.038 & 0.2 & 1.0%
\\ \hline
\end{tabular}
\caption{{\small{The stable isotopes of the chemical elements that compose the  
DNA/RNA macromolecules and their relative abundances, in percentages. Values from the URL 
{\color{blue}{\url{http://www.periodictable.com/Isotopes/001.1/index.dm.html}}}.}}}
\label{dna}
\end{table}
\begin{figure}[tbhp]
\includegraphics[height=1.8in, width=2.5in]{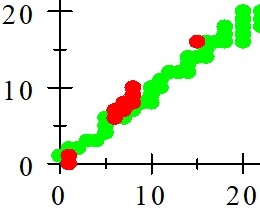}
\caption{\small{The red dots (or as dark gray for an image in fifty shades of gray) correspond 
to the five elements composing, in a quite particular structure, the DNA/RNA macromolecules. 
This figure is a blow up of the lower part of Fig. \ref{figestaveis}.}}
\label{figdna}
\end{figure}
So, it is sensible to conjecture that if one or more of the five chemical elements within the DNA/RNA 
did contain a natural radioactive isotope (\emph{i.e.} one with a non-null isotopic abundance), it 
would be embodied within the DNA/RNA, so turning unstable sectors of these macromolecules. After decaying 
through $\beta$ emission, the molecular structure would be eventually destroyed or it would induce a 
transcription with errors (or even its impossibility) for the synthesis of proteins. For example, if the 
tritium ($^{3}H$), an hydrogen isotope, had on Earth a non-null relative abundance\footnote{Tritium is 
produced in nuclear reactors by, for instance, thermal neutron activation of lithium-6: 
%
\[
n+\ ^6_3Li \longrightarrow \ ^4_2He \ (2.05\ MeV)	+ \ ^3_1H \ (2.75\ MeV); 
\]
it is an exothermic reaction yielding an energy of $4.8\ MeV $.}, it would be present in the DNA/RNA 
macromolecule in the same proportion. Still, as its half-life time is about twelve years, it decays 
and transmutes into a helium-3 atom which is stable and has a quite small relative abundance as 
compared to the other isotope, $^4He$: $1:730.000$. Nevertheless, due to its specific electronic 
structure it is an inert atom, that would not bind chemically with another element to make a stable 
electrically neutral molecule\footnote{Nevertheless the helium hydride exists as a compound cation, it 
was produced in laboratory more than one hundred years ago. Recently, its presence was observed in the 
outer space and it is considered as the primordial molecule in the Universe \cite{HeH}.}, thus it would 
originate a different DNA structure whose role could be quite distinct from that of the current DNA/RNA. 
In short, one could cogitate -- actually a truism -- that Nature chose criteriously the suitable chemical 
elements to build the molecular structures that conceal the informational content which codifies the 
proteins. In Fig. \ref{figdna} we present part of the valley of stability where, highlighted in red color,
 we find the locations of the nuclides constituting the DNA/RNA. 
%
\section{The mass formula frontier lines and the chart of nuclides delimiting lines}\label{linhas}
%
In subsection \ref{estab} we have seen that the calculated line of stability fits quite neatly the 
distribution of the stable $252$ nuclides, that constitute the valley of stability, whose half-life 
times are assumed to be ``infinite''. In this section we address the question: what else the mass formula 
can reveal about the most unstable radionuclides by still using the same set of parameters from 
\cite{WAPSTRA1958,WAPSTRA1985}. A striking fact about natural radionuclides, 
of the same chemical element, is that their half-life times may vary, from one another, by several orders of 
magnitude. For example, the lead, which has the highest atomic number in the valley of stability, $Z=82$, 
has four stable isotopes, $^{204}Pb,\ ^{206}Pb,\ ^{207}Pb,\ ^{208}Pb$, with finite relative abundance and, 
in addition, there are about 34 radioisotopes, with null relative abundances, that originate from the 
decays of other radionuclides. Those having the highest excess of protons and neutrons, are the $^{178}Pb$, 
with half-life time $T_{1/2} \approx 230\ \mu s $, and the $^{215}Pb$ with $T_{1/2} \approx 36\ s$, respectively, 
both belong to the set of nuclides that form the matter delimiting lines. Among the chemical elements that have 
no isotopes in the valley of stability we may consider, for instance, the natural element thorium, $Z=90$, 
observing that while the $^{232}Th$ isotope has a relatively long half-life time, $T_{1/2} \approx 14\ Gy$ 
(the same order of our Universe estimated age) and $100\%$ relative abundance it is not considered being a 
stable nuclide. Thorium has about 30 isotopes and among them the $^{209}Th$ has a half-life time $T_{1/2} 
\approx 3.8\ ms $ and the $^{238}Th$ with $T_{1/2} \approx 9.33\ m $, both belonging to the delimiting lines. 
Henceforth our aim here is to still use appropriately the mass formula in order to: (a) draw frontier lines 
where, between and on them, one can localize most of the nuclides, radioactive or stable, and (b) estimate at 
what extent the frontier lines coincide with the delimiting lines as presented in the chart, Fig. \ref{carta}. 
The results should allow us to evaluate the power and the versatility of the mass formula. 
 
The association of the mass formula, Eq. (\ref{massa1}), with the \emph{energy-time} uncertainty relation, 
which is a fundamental concept of quantum mechanics, can widen the application scope of the original classical 
drop model. In what follows we show how we can use that relation in order to determine the protons and neutrons 
frontier lines. As in quantum mechanics time is a parameter and not an operator, that relation can be explored 
quantitatively insofar as the precise sense of the involved physical quantities are worked out adequately 
\cite{pauli}. In references \cite{PIZA2002,messiah,griffiths,cohen}, for instance, more detailed discussions 
can be found on how that relation is proposed and justified, and under which conditions it can be used.

We now relate the \emph{uncertainty in energy} to a specific physical process: the energy associated 
with the transformation of a nucleus $_Z^AX$ into another one with same $A$, but with one less proton or 
one less neutron, $_{Z \mp 1}^{\; \; A}X$, i.e., processes through which a change of only one coulomb 
charge unit occurs within the nucleus. By its turn, the \emph{uncertainty in time} $\Delta t$ should be 
associated with that specific transformation. In order to establish an expression for $\Delta t$ we need 
to determine typical time scales for such processes, so a semi-classical approach can guide us onward this 
direction. 

In the chart in Fig. (\ref{carta}) we observe that, as far as the current experimental results indicate, 
at the two extreme points along an isobar line, there is a nuclide that has the greatest possible 
proton number (relatively to the neutron number) and another one with the greatest neutron number (relatively 
to the proton number); the set of those nuclides for all $A$'s, constitute, as already mentioned, the 
delimiting line of protons or neutrons, respectively. Those radionuclides are supposed to be the most 
unstable and it is admitted that the delimiting lines change their line trace whenever new nuclides are produced 
artificially, discovered in astrophysical bodies or observed in the outer space.

We start by calculating the nucleus energy difference when, for a given fixed $A$, the proton number $Z$ is 
augmented by one unit through some process as, for example, in a reaction of the kind 
\be
p+\ _{Z-1}^{\; \;A}X\rightarrow \ _{\; \; Z}^{A+1}Y^{\ast }\rightarrow n+\ _{Z}^{A}Y\ . 
\ee
which is different from a $\beta$ decay due to the disparity of the time scales. As will be discussed ahead, 
one imposes that the intermediary nuclear structure $\ _{\; \; Z}^{A+1}Y^{\ast }$ must exist for a ''very short'' 
time interval ($\gtrsim 10^{-23}\ s$) and the adopted associated energy (mass) difference is 
$\Delta M\left( A,Z\right) =M\left( A,Z\right) -M\left( A,Z-1\right) $, which, in terms of the coefficients 
of Eq. (\ref{masscoef}), is then given by 
\begin{equation}
\Delta M\left( A,Z\right) =C_{2}-C_{3}\left( A\right) +2C_{3}\left( A\right)
Z\  ,
\label{eq3}
\end{equation}
observing that this expression is linear in $Z$. In fact, it consists in the calculation of the 
energy difference along one isobar line, as can be promptly observed from the IAEA chart \cite{IAEA}. 
The pairing term (\ref{eta}) that is part of the mass formula does not contribute to Eq. (\ref{eq3}) 
because of a cancellation of two terms that are equal in modulus but have opposite signals. 
\begin{figure}[htb]
\centering
\includegraphics[height=3.7in, width=4.9in]{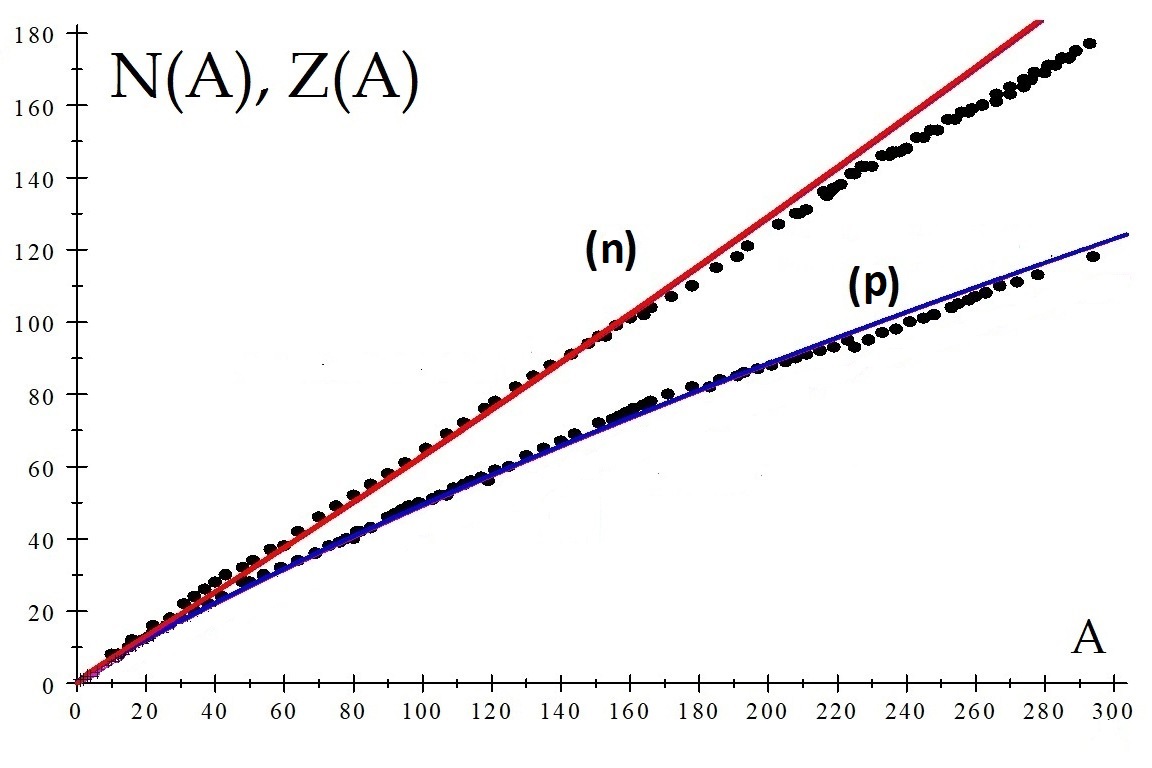}
\caption{\small {The dots in black correspond to two sets of nuclides picked from the left and 
right borders of the chart as seen in Fig. \ref{carta}, each set characterizes a \emph{delimiting line}. 
The label (n) is for the neutrons and (p) is for the protons. The solid lines are the 
\emph{frontier lines} as calculated from the mass formula.}}
\label{linhas2}
\end{figure}

The increment of one charge unit in the nucleus $_{Z-1}^{\; \;A}X$ changes it into another one, 
$_{Z}^{A}Y$, which becomes unstable and promptly emits a nucleon or decays by another energetically 
allowed process, as displayed in Eq. (\ref{addp}). This decay will guide us in establishing the 
magnitude of $\Delta t$ that will set a limit on the number of nucleons that defines the frontier 
lines; i.e., the time interval must be associated with the instability of the nucleus that, in 
principle, should hamper the existence of nuclides beyond these lines. We are not seeking details of 
the nuclear processes within the nuclei, but showing that by only using fundamental physical concepts 
we can set a condition for not allowing the existence of nuclides beyond the frontier lines, also 
referred as drip lines, that delimit the region away from which a nucleus promptly decays emitting 
a proton or a neutron. Now, we elaborate more on $\Delta t$ for the case of one more proton inside the 
nucleus. Starting from the relation 
\be
\Delta M\left( A,Z\right) c^{2}\Delta t\sim \hbar \ , 
\label{deltaM}
\ee
one notes that the quantum theory manifests itself explicitly through the presence of Planck's constant. 
As an estimate of the lower limit for $\Delta t$ we first consider it being the time spent for a 
free nucleon to cross a distance that is a typical diameter of a nucleus, having the speed of light in 
vacuum. Furthermore, resorting to the drop model the nuclear radius is $R=r_{0}A^{1/3}$, 
where $r_{0}=1.2\ fm$ ($1\ fm=10^{-15}\ m$) is a typical parameter adjusted in order to give, in 
average, the radius of a nucleus having mass number $A$ \cite{MG2016}. For a nucleus with a $10\ fm$ 
diameter the order of magnitude of that time is $\Delta t={2R}/{c}\sim 10^{-23}\ s$. Under these 
conditions this is the lowest time interval a nucleon can exist inside a nucleus. A more realistic 
$\Delta t$ value for the incident proton implies considering a velocity $v<c$ in the system center of 
mass reference frame. Thus, with $\beta = v/c$ we write
\be
\Delta t\left( A\right)  =\frac{2r_{0}A^{1/3}}{\beta c} ,
\ee
that depends on the mass number of the nucleus. Therefore, we get the relation 
\be
\Delta M\left( A,Z\right) \left[ MeV \right] =  \frac{\hbar }{\Delta t\left( A\right) }%
\simeq 82.22\ \beta A^{-1/3} \ ,
\ee
that allows us to write an equation for $Z$ as a function of $A$, 
\begin{equation}
Z\left( A\right) = \frac{\hbar }{2C_{3}\left( A\right)
c^{2}\Delta t\left( A\right) }-\frac{C_{2}-C_{3}\left( A\right) }{%
2C_{3}\left( A\right) } 
\label{eq4}
\end{equation}
which determines the frontier line for the nuclides having the highest proton excess. In order 
to compare $Z\left( A\right)$ with the dots that define the delimiting line of the chart one 
could have considered the integer $A$ and the closest integer that results from the calculation 
of the RHS of Eq. (\ref{eq4}), namely Int$[Z\left( A\right)]$. However, we preferred to do the 
comparison drawing a continuous line because, visually, it facilitates the analysis. We chose 
$\beta = 0.7$ not with the purpose to get the best fit to the data but to simply represent an 
admissible value for $\Delta t$. In fact, we verified that other values, close to $0.7$, do not 
change substantially the results. The essential feature of $\Delta t\left( A\right)$ is its 
dependence on $A^{1/3}$. Performing the comparison, as presented in Fig. \ref{linhas2}, one notes 
that although the calculated frontier line (p) does not coincide exactly with the delimiting line 
(black dots), the overall agreement is quite satisfactory. We also observe a fair agreement even 
in the region of low values of $A$, where the liquid drop description of nuclei are usually 
admitted to be poor. 

To reduce the observed deviations one can proceed by: (a) adding to the mass formula more terms 
containing additional empirical parameters \cite{SEEGER1961,myers,moller} or, (b) embracing a 
different approach that consists in using a microscopic model to calculate for each nuclide its 
nuclear properties, choosing, for instance, a Hartree-Fock shell model altogether with many-parameter 
phenomenological two and three-body forces; in this case the resulting values incorporate the quantum 
many-body effects, see Refs. \cite{RING1980,ROWE2010}. Although the procedure (b) shall lead to more 
detailed and precise description of a nuclide properties, it asks for a quite arduous calculation effort, 
along with a significant demand of computational resources in terms of time consumption and memory storage, 
thus putting this approach out of the scope of our current proposal. Concerning procedure (a), more terms 
means more free parameters to be introduced, that, although improving the agreement with the data, will  
devoid the mass formula \ref{massa1} from its fundamental simplicity \cite{WEIZS1935,BETHE1936}.

In the same form as we did for the protons we now proceed with the calculation for the neutron frontier 
line, where the energy difference is
\be
\Delta M\left( A,Z\right) =M\left( A,N\right) -M\left( A,N-1\right) 
\ee
and, here too, making use of the energy-time uncertainty relation we get an equation for 
those nuclides of mass number $A$ that have the highest neutron excess, 
\be
N\left( A\right) = \frac{1}{D_{3}\left( A\right) }\left[
D_{1}\left( A\right) + D_{2}\left( A\right) +\frac{\hbar }{\Delta t\left(
A\right) }+\left( M_{n}-M_{1_{H}}\right) \right]  , 
\label{eq5}
\ee
where
\be
\left.
\begin{array}{lll}
D_{1}\left( A\right) &=& 4a_{ass}-a_{c}A^{-1/3}  \\ 
D_{2}\left( A\right) &=& 4a_{ass}A^{-1} + 2a_{c}A^{2/3} \\
D_{3}\left( A\right) &=& 2a_{c}A^{-1/3}+8a_{ass}A^{-1} \ ,
\end{array}
\right\}
\label{masscoefn}
\ee
that defines the neutron frontier line. In Fig. \ref{linhas2} it corresponds to the solid line 
denoted by (n); we used the same set of parameters and $\beta =0.7$. From the chart one gets 
the set of nuclides having the maximum neutron excess, for all the isobar lines, that makes the 
neutron delimiting line represented by the black dots. Alternatively, we could also have compared 
the set of the closest integers that result from the calculation of the RHS of Eq. (\ref{eq5}), 
namely Int$[N\left( A\right)]$, with those dots, but as said above, here too, we preferred to do the 
comparison using the frontier (solid) line. In this neutron excess case the agreement is also quite 
good up to the nuclides with mass numbers $A \approx 180$; only above this number a pronounced 
deviation appears and it could also be attributed to the descriptive limitation of the drop model 
for nuclear systems, consequently, we may admit that the introduction of quantum many-body effects 
hints to be essential for a refined adjustment. 

Nevertheless, we conjecture that this deviation could be partially attributed to the lack of nuclides, 
those that still have not been produced artificially, as pointed, for instance, in an experiment 
reported in \cite{BAUMANN2007}, although in this paper the nuclides are in the region $A \approx 40$. 
The production/discovery of new nuclides with higher neutrons excess would shift the delimiting line 
trace (linking the dots) toward the frontier line, thus reducing the distance that separates them. In 
this case the remaining deviations could then be attributed to the quantum many-body effects not present 
in the mass formula. 

In the quest for improving the agreement between the frontier lines with the phenomenological 
delimiting lines one pertinent question to be answered is: can the naive expression for the nuclear 
radius $R(A) =r_{0}A^{1/3}$ be the object of further additional phenomenological changes, for 
example, by introducing explicitly in it the proton and neutron numbers $Z$ and $N$? For instance, 
by taking into account the difference $(N-Z)$? This redefinition of the nuclear radius is inspired 
by protons or neutrons excess that, in principle, could be at the origin of a nuclear halo out of a 
central part occupied by equal numbers of protons and neutrons \cite{LIU2004,KOBA2014}. Such a form, 
proposed in \cite{turcos}, is  
\be
R\left( A; Z,N \right) =r_{0}\left( 1-b\left({N-Z}\right)/{A}\right) A^{1/3} , 
\label{rturcos}
\ee
with the numerical values $r_{0}=1.269\ fm$ and $b = 0.252$ obtained from the nuclear radii best 
fit using a large number of experimental data. We carried out the calculations for protons and neutrons 
frontier lines using Eq. (\ref{rturcos}) and still keeping $\beta =0.7$. The results were compared 
with those previously obtained and we observed that there is no appreciable change in the behavior 
of the frontier lines, they practically coincide with those calculated using $R(A)=r_0 A^{1/3}$. This 
fact corroborates our guess that for determining the frontier lines the dependence on $A^{1/3}$ is the 
essential element to describe the delimiting lines, thus dispensing the necessity to introduce further 
parameters in $\Delta t(A)$. 

As an extra observation, we find out that the angle between the two frontier lines in Fig. \ref{linhas2} 
is approximately $7.5^\circ$, and within this narrow region one finds all the known matter constituents 
(different kinds of nuclides), as predicted by the drop model and substantiated by the empirical data. 
The ratio of this angle to the $90^\circ$ quadrant shows that the distribution of the possible nuclides 
occupies nearly $8.3\%$ of it. 
%
\section{Summary and conclusions}\label{conc}
%
Starting from basic ideas and concepts of nuclear physics (nuclide mass formula) and quantum 
mechanics (energy-time uncertainty relation) we obtained a quantitative evaluation of the 
frontier lines of matter, which we compared with the experimental delimiting lines just as 
they show up in the chart of nuclides. As a matter of fact, the approach is an extension of 
a clue presented in the authors textbook \cite{MG2016} that was not pursued more thoroughly 
at that time, a project which we have implemented here. Our approach illustrates the strength 
of the energy-time uncertainty relation that was fundamental to determine the frontier lines 
associated with the existence of highly unstable nuclei -- a condition that characterizes a 
nucleus having an excess in its number of protons over its neutrons and \emph{vice versa} -- 
keeping their identity during the minimal time interval necessary to measure some of their 
properties. The adopted plan was: (a) to calculate, from the mass formula, the difference in 
the energy associated with the process by which a nucleus coulomb charge is changed by one unit 
and still keeping the same mass number $A$; and (b) to estimate the average time interval during 
which the nucleus remains in such an unstable state, that was assumed being the ratio of the 
``nuclear diameter'' of a target nucleus to the speed of a nucleon hitting it and forming a 
compound system prior to its disintegration. 

In this way, our calculation, based on consolidated nuclear physics arguments, purveys a description 
consistent with the empirical data as presented in the chart of nuclides, and it is important 
to underline the fact that there was no need to introduce new, or modified, parameters for obtaining 
the frontier lines. Concerning the line associated with the neutron excess (n), we conceive that 
the observed deviations from the experimental data for $A>180$, (compared with the line (p) of proton 
excess) are partially due to the absence of quantum many-body effects. Still, we cannot rule-out 
the possibility that the production and discovery of new nuclides having higher neutron excess, 
would shift the dots of the delimiting line moving it closer to the frontier line (n), thus 
reducing their separation gap, and hence making their agreement more similar to that one observed 
between the proton lines (p). Even so, in the case that this possibility is not likely to be 
successfully ratified by experimental quests, the confirmation that our approach, much less than 
a refined theory, allowed an almost precise delineation of the matter delimiting lines is a credit 
for our heuristic formulation that links the nuclide mass formula to the energy-time uncertainty 
relation of quantum theory. 

In conclusion, we remark that we have found useful to also present to the reader a broader introduction 
to the nuclear physics essentials and a brief narrative of the main trend of the history of 
the nucleosynthesis and the origin of matter in the Universe. We complemented the paper with an aside 
topic linked to nuclides: a discussion about the nature of the chemical elements that compose the DNA/RNA 
macromolecules and the fact that none of them has a natural radioactive isotope. 
%
\section*{Acknowledgements}
%
{SSM thanks the Conselho Nacional de Desenvolvimento Científico e Tecnológico, CNPq, for financial support 
as a research grant. The authors are grateful to Dr. Marcelo A. Marchiolli for valuable suggestions.}
\newpage
\end{document}